# A simple Python code for computing effective properties of 2D and 3D representative volume element under periodic boundary conditions

Fan Ye, Hu Wang[*]

State Key Laboratory of Advanced Design and Manufacturing for Vehicle Body, Hunan University, Changsha 410082, P.R. China

**Abstract** Multiscale optimization is an attractive research field recently. For the most of optimization tools, design parameters should be updated during a close loop. Therefore, a simple Python code is programmed to obtain effective properties of Representative Volume Element (RVE) under Periodic Boundary Conditions (PBCs). It can compute the mechanical properties of a composite with a periodic structure, in two or three dimensions. The computation method is based on the Asymptotic Homogenization Theory (AHT). With simple modifications, the basic Python code may be extended to the computation of the effective properties of more complex microstructure. Moreover, the code provides a convenient platform upon the optimization for the material and geometric composite design. The user may experiment with various algorithms and tackle a wide range of problems. To verify the effectiveness and reliability of the code, a three-dimensional case is employed to illuminate the code. Finally numerical results obtained by the code agree well with the available theoretical and experimental results.

**Keywords** Representative Volume Element  Periodic Boundary Conditions  Asymptotic Homogenization Theory  Multiscale

## 1. Introduction

Prediction of the mechanical properties of the composites has become an active research area [1-3]. Except for experimental studies and macromechanical methods, micromechanical methods are widely used to obtain overall properties of composites. The methods usually contain the introduction of two scales, marcroscale and microscale. Marcroscale is usually referred to a homogenized continuous medium,

[*] Corresponding author, wanghu@hnu.edu.cn, Tel: +86-0731-88821417, Fax: +86-0731-88822051

and microscale is usually related to a statistically RVE [4]. The RVE must be selected such that the microstructure should be composed as copies of RVEs, and without overlapping boundaries or gaps between boundaries. The choice of RVE is usually not unique, but it should be big enough to present the feature of composite and as small as possible to reduce the computational cost. Additionally, the RVE should have the same volume fraction as the composite [5]. Micromechanical methods provide overall behaviors of the composites from known properties of their constituents (inclusion and matrix) through an analysis of a RVE [6-7].

The above micromechanical models can be regarded as mechanical or engineering models. AHT is proposed to predict overall behaviors of micromechanical models [8-11]. Moreover, the AHT has explicitly used periodic boundary conditions in modeling of linear and nonlinear composite materials [12]. On the other hand, the theory for predicting overall behaviors of micromechanical models is laid on a rigorous mechanical foundation by using strain energy equivalence principles in conjunction with Finite Element (FE) analysis [13]. Explicit unified form of boundary conditions for RVE is developed which satisfies the periodicity conditions, and is suitable for any combination of multiaxial loads [14].

The AHT combined with the FE method is proven to be a powerful technique [15-16], which can consider a more complex microstructure given by several inclusions with different shapes, different orientations and different aspect ratios, and even random distribution of inclusions. For this theory based on FEM, there is no restriction on geometry and material, and the AHT based FE has been well documented for the determination of the effective material properties of the composites.

ABAQUS is a general purpose commercial software package widely used in the analysis of the RVEs [17, 18]. Moreover, the Python code takes advantage of the advanced FE analysis capacities of ABAQUS software and employs the ABAQUS Scripting Interface (ASI) to communicate with ABAQUS. ASI is an extension of the Python language. As a standard component of the ABAQUS software, ASI provides a convenient interface to the models and results [19]. In this study, a Python code is developed for computing effective properties of the RVE under PBCs based on the secondary development of ABAQUS.

## 2. Periodic boundary conditions

Assumed that the composite is heterogeneous with a periodic microstructure, the RVE is subjected to boundary conditions depending on macroscale. The displacement field for the periodic structure can be expressed as [8]

$$u_i(x_1, x_2, x_3) = \bar{\varepsilon}_{ik} x_k + u_i^*(x_1, x_2, x_3) \tag{1}$$

where $\bar{\varepsilon}_{ik}$ is the macro (average) strain tensor of the periodic structure and the first term on the right side represents a linear distributed displacement field. The second term on the right side, $u_i^*(x_1, x_2, x_3)$, is a periodic function from one RVE to another. It represents a modification to the linear displacement field due to the heterogeneous structure of the composite.

Since the periodic array of RVEs represents a continuous physical body, two continuities must be satisfied at the boundaries of neighboring RVEs. One is that displacements must be continuous, and neighboring RVEs cannot be separated or overlap each other at the boundaries after deformation. The second condition implies that the traction distributions at opposite parallel boundaries of a RVE must be the same. In this manner, the individual RVE can be assembled as a physically continuous body.

Obviously, the assumption of the displacement field in the form of Eq. (2) meets the first of above requirements. Unfortunately, it cannot be directly applied to the boundaries since the second term on the right side of Eq. (1) is generally unknown. For any RVE, its boundary surfaces must appear in parallel pairs, the displacements on a pair of parallel opposite boundary surfaces can be written as

$$u_i^{j+} = \bar{\varepsilon}_{ik} x_k^{j+} + u_i^* \tag{2}$$

$$u_i^{j-} = \bar{\varepsilon}_{ik} x_k^{j-} + u_i^* \tag{3}$$

where indices on the left side identify the pair of two opposite parallel boundary surfaces of a RVE.

It should be noted that $u_i^*(x_1, x_2, x_3)$ is the same at two parallel boundaries (periodicity), therefore, the difference between above two equations is

$$u_i^{j+} - u_i^{j-} = \bar{\varepsilon}_{ik}(x_k^{j+} - x_k^{j-}) = \bar{\varepsilon}_{ik} \Delta x_k^j . \tag{4}$$

Since $\Delta x_k^j$ are constants for each pair of the parallel boundary surfaces, with specified $\bar{\varepsilon}_{ik}$, the right side becomes constants. Such equations can be easily applied to FE analysis as nodal displacement constraint [14]. Moreover, if RVE is analyzed by using a displacement-based FE method, the application of Eq. (4) can guarantee the uniqueness of the solution and traction continuity conditions are automatically satisfied instead of applying Eq. (1) directly as the boundary conditions.

## 3. Homogenization method

In the homogenization method, the RVE is modeled as a homogeneous orthotropic medium with certain effective properties that describe the average material properties of the composite [11]. To describe this macroscopically homogeneous medium, macrostress and macrostrain are derived by averaging the stress and strain tensor over the volume of the RVE:

$$\bar{\sigma}_{ij} = \frac{1}{V_{RVE}} \int_V \sigma_{ij}(x,y,z) dV \quad (5)$$

$$\bar{\varepsilon}_{ij} = \frac{1}{V_{RVE}} \int_V \varepsilon_{ij}(x,y,z) dV \quad (6)$$

The strain energy $U^*$ stored in the heterogeneous RVE of the volume $V_{RVE}$ is

$$U^* = \frac{1}{2} \int_{V_{RVE}} \sigma_{ij} \varepsilon_{ij} dV \quad (7)$$

Instead the strain energy corresponding to the heterogeneous RVE, the homogeneous RVE is used to determine the corresponding homogenized modulus. In the homogeneous RVE, the total strain energy due to deformation is given by

$$U = \frac{1}{2} \bar{\sigma}_{ij} \bar{\varepsilon}_{ij} V_{RVE} \quad (8)$$

where $U$ is homogeneous strain energy of RVE. $\bar{\varepsilon}_{ij}$ is average strains. $V_{RVE}$ is volume of periodic domain. $\bar{\sigma}_{ij}$ is average stresses.

As shown in [13], Eq. (7) and Eq. (8) can be converted to a surface integral by using Gauss theorem, thus:

$$U^* - U = \frac{1}{2} \int_{S_j} \sigma_{ij}(u_i - \bar{u}_i) n_j dS_j \quad (9)$$

where $S_j$ is the *j*th surface and $n_j$ is the unit outward normal. $u_i$ is the *i*th displacment. $\bar{u}_i$ is the *i*th average displacment. On the surface $S_j$:

$$u_i = \bar{u}_i \quad (10)$$

Thus:

$$U^* - U = 0 \tag{11}$$

The average stresses and strains quantities defined in Eq. (5) and Eq. (6) ensure equivalence in strain energy between the equivalent homogeneous material and the original heterogeneous material.

Alternatively, the average strains can be related to boundary displacements of the RVE by using Gauss theorem. Average strains in Eq. (6) can be converted as:

$$\bar{\varepsilon}_{ij} = \frac{1}{V_{RVE}} \int_{S_1} (u_i n_j + u_j n_i) dS_1 \tag{12}$$

where $S_1$ denotes the outer boundary of the RVE. The relationship given by Eq. (12) makes it possible to evaluate the volume averaged strains with using the boundary displacements, and then avoiding the volume integration [13].

The average stresses $\bar{\sigma}_{ij}$ is obtained by the prescribed average strains $\bar{\varepsilon}_{ij}$. The tensors are the ratio of average stresses and strains and computed as follow:

$$C_{ij} = \frac{\bar{\sigma}_{ij}}{\bar{\varepsilon}_{ij}} \tag{13}$$

where $C_{ij}$ is stiffness modulus corresponding to apply deformation mode.

In the case of pure shear deformation, the tensorial shear strain $\varepsilon_{ij}$

$$\gamma_{ij} = \varepsilon_{ij} + \varepsilon_{ji} = 2\varepsilon_{ij} \tag{14}$$

where $\gamma_{ij}$ is the engineering shear strain. In this study, the tensorial shear strains are suited to be the macro strain [20].

## 4. A three-dimensional case

To verify the homogenization method under PBCs, a 3-D RVE model is considered. The model consists of a fiber reinforcement and matrix, with a volume fraction of 47 %. The properties of the materials are given in Table 1.

Table 1 Material properties of fiber and matrix

| Material | E (Gpa) | v |
| --- | --- | --- |
| Fiber | 379.3 | 0.1 |
| Matrix | 68.3 | 0.3 |

The unidirectional RVE is assumed to be orthotropic and linearly elastic. For 3D RVE, from Eq. (13), the material constitutive relation of this RVE can be written as

$$[\bar{\sigma}] = [C][\bar{\varepsilon}] \tag{15}$$

where $\bar{\varepsilon}$ is the average strain matrix. $\bar{\sigma}$ is the average stress matrix. $[C]$ is the stiffness matrix as shown in Eq. (16) for the orthotropic and unidirectional RVE.

$$\begin{bmatrix} \bar{\sigma}_{11} \\ \bar{\sigma}_{22} \\ \bar{\sigma}_{33} \\ \bar{\sigma}_{12} \\ \bar{\sigma}_{13} \\ \bar{\sigma}_{23} \end{bmatrix} = \begin{bmatrix} C_{11} & C_{12} & C_{13} & 0 & 0 & 0 \\ C_{12} & C_{22} & C_{23} & 0 & 0 & 0 \\ C_{13} & C_{23} & C_{33} & 0 & 0 & 0 \\ 0 & 0 & 0 & C_{44} & 0 & 0 \\ 0 & 0 & 0 & 0 & C_{55} & 0 \\ 0 & 0 & 0 & 0 & 0 & C_{66} \end{bmatrix} \begin{bmatrix} \bar{\varepsilon}_{11} \\ \bar{\varepsilon}_{22} \\ \bar{\varepsilon}_{33} \\ \bar{\varepsilon}_{12} \\ \bar{\varepsilon}_{13} \\ \bar{\varepsilon}_{23} \end{bmatrix} \quad (16)$$

After obtaining $\bar{\sigma}_{ij}$ for given $\bar{\varepsilon}_{ij}$ from the computation of a RVE, $C_{ij}$ can be obtained from Eq. (15). From Eq. (8) and Eq. (15), the relation between the homogeneous strain energy and the average strain is

$$U = \frac{1}{2}[\bar{\varepsilon}]^{\mathrm{T}}[\bar{\varepsilon}]V_{\mathrm{RVE}} \quad (17)$$

Thus

$$\begin{aligned} \frac{U}{V_{\mathrm{RVE}}} &= \frac{1}{2}C_{11}\bar{\varepsilon}_1^2 + C_{12}\bar{\varepsilon}_1\bar{\varepsilon}_2 + C_{13}\bar{\varepsilon}_1\bar{\varepsilon}_3 + \frac{1}{2}C_{22}\bar{\varepsilon}_2^2 + C_{23}\bar{\varepsilon}_2\bar{\varepsilon}_3 \\ &+ \frac{1}{2}C_{33}\bar{\varepsilon}_3^2 + \frac{1}{2}C_{44}\bar{\varepsilon}_4^2 + \frac{1}{2}C_{55}\bar{\varepsilon}_5^2 + \frac{1}{2}C_{66}\bar{\varepsilon}_6^2 \end{aligned} \quad (18)$$

Hence, the values of the stiffness matrix are computed from Eq. (18).

## 5. Python implement of the three-dimensional case

This section explains the basic form of the Python codes. The code is to be called in ABAQUS/CAE through the menu command File-> Run Scipt. A flowchart of the code is presented in Fig.1.

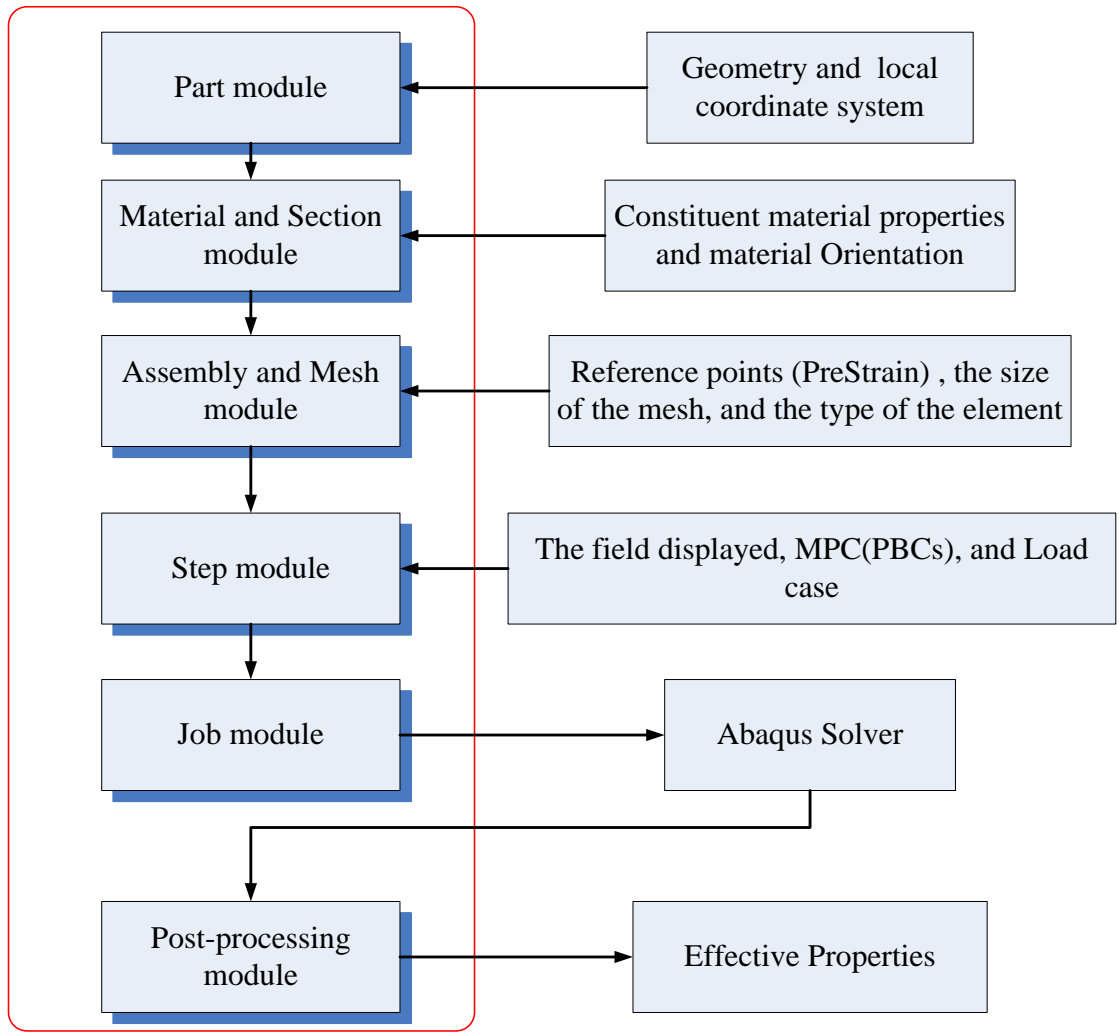

Fig. 1 The work flowchart of the Python code for the computation of RVE

## 5.1. Header

As a common programming practice, a header exists at the beginning of the script to import a number of modules and functions. For the four modules, *abaqus*, *abaqusConstants*, *caeModules*, and *odbAccess*, they are standard components of the ABAQUS software to employ the ASI. The *numpy* and *math* modules are used to import arithmetic functions.

```
from abaqus import *
from abaqusConstants import *
from caeModules import *
from odbAccess import *
from numpy import *
import math
```

## 5.2. Geometric parameters of the RVE (Create part)

The shape of the RVE is decided by the microstructure, which can be an accurate reflection of reality or an idealisation. While it remains at the user's discretion which idealisation is adopted, considerations should be made so as to have the idealised microstructure as representative of reality as possible. In this study, in the case of uniderectionally fiber-reinforced composites with fibers distributed at the center of the RVE, the square packing is selected to be idealistic representive element of the composites.

This function runs the ABAQUS to creat the shape of the RVE. The geometric parameters of the RVE are defined into $W_x = W_y = W_z = W = d = 1mm$ as shown in Fig.2. Moreover, the fiber volume fraction of the model is designed to be 47%.

```
##Function of creating RVE part
def CreatePart(RVE_para,RVEModel ):
    #Initial parameter
    # Create  RVE sketchs
    RVEModel.ConstrainedSketch(name='', sheetSize='')
     #Define two point to creat a rectangle
    RVESketch.rectangle(point1='', point2='')
    #Define the property of the part
    RVEPart= RVEModel.Part(name='',dimensionality='',type='')
     #Assign sketch to the part
    RVEPart.BaseSolidExtrude(RVESketch, depth=)
     #Create fiber sketchs
    RVEt = RVEPart.MakeSketchTransform(sketchPlane='', sketchUpEdge='')
    RVESketch1= RVEModel.ConstrainedSketch(name='', sheetSize=,transform=)
    #Define two point to creat a Circle
    RVESketch1.CircleByCenterPerimeter(center=, point1=)
    #Assign fiber sketch to the part
     RVEPart.PartitionFaceBySketch(sketchUpEdge=,faces=, sketch=)
    #Partition the fiber
    RVEPart.PartitionCellByExtrudeEdge(line=,cells=,edges=,sense=)
    return RVEPart
```

## 5.3. Definition of constituent materials (Create materials and section)

This function allows for the definition of each phase of the composite, inclusion/reinforcement and matrix. In this study, two material types are available for the composite. The matrix and inclusion are each defined in a local coordinate system to account for the material orientations as appropriate. The material properties to be entered are in terms of engineering constants, Young's modulus and Poisson'ratio for elastic behaviours. The material properties of the RVE are given in Table 1. Moreover, the principal direction of the RVE is *z*-direction.

```
##Function of creating materials for RVE
def CreateMaterial(RVE_para,RVEModel ):
    #Define the name of two materials
    RVEModel.Material(name='')
    RVEModel.Material(name='')
    #Define the property of  materials
    RVEModel.materials[''].Elastic(table=)
    RVEModel.materials[''].Elastic(table=)
##Function of creating sections for RVE
def CreateSection(RVE_para,RVEModel ,RVEPart):
    #Initial parameter
    #Define the name of the Section
    RVEModel.HomogeneousSolidSection(name='',material='',thickness=)
    RVEModel.HomogeneousSolidSection(name='',material='',thickness=)
    #find the face
    #Assign the property of the section1
    RVEPart.SectionAssignment(region=, sectionName='')
    RVEPart.MaterialOrientation(region=, orientationType=, axis=,additionalRotationType=,
                    localCsys=,fieldName='', stackDirection=)
    #find the other face
    #Assign the property of the section2
    RVEPart.SectionAssignment(region=, sectionName='')
    RVEPart.MaterialOrientation(region=, orientationType=, axis=,additionalRotationType=,
                    localCsys=,fieldName='', stackDirection=)
```

## 5.4. Key degrees of freedom and meshing (Create assembly)

Macroscopix/average strains $(\varepsilon_x^0, \varepsilon_y^0, \varepsilon_z^0, \varepsilon_{xy}^0, \varepsilon_{xz}^0, \varepsilon_{yz}^0)$ appearing in boundary conditions are physical entities which can be assigned by independent node numbers and treated as ordinary nodes or Degrees of Freedoms (DOFs). The nodal 'displacements' (dimensionless) at these special DOFs give the corresponding macroscopic strains directly, and eliminate the need to obtain them by averaging strains from all elements. The macroscopic/average strains $(\varepsilon_x^0, \varepsilon_y^0, \varepsilon_z^0, \varepsilon_{xy}^0, \varepsilon_{xz}^0, \varepsilon_{yz}^0)$ can be assigned to the 'Boundary constraints' which are related to the boundary displacement of the RVE.

In the terminology of ABAQUS, these key DOFs have been introduced as 'reference points'. The key DOFs representing the macroscopic strains appear in the PBCs that they become physical entities as part of the RVE.

The FE analysis requires reasonable meshes for numerical convergence considerations. For RVEs, extra restrictions must be imposed on meshes. To impose periodic boundary conditions on a RVE, the paired faces corresponding to periodic boundary conditions must be tessellated identically. This can usually be achieved by copying one tessellated surface to another before generation of mesh inside the RVE between paired faces. The mesh of the RVE may be appropriately formulated based on rational considerations of boundary symmetries. In this study, the infusion is at the center of the RVE, and it can be easy to keep the mesh of the faces periodicity through the Python script without user involvement.

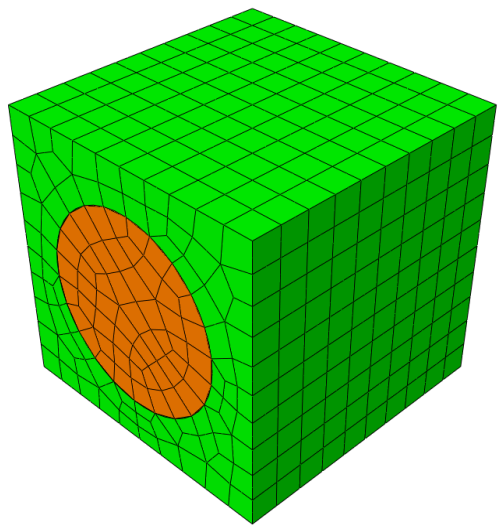

Fig.2 FE mesh of RVE model for the unidirectional composite

```
##Function of creating assembly for RVE
def CreateAssembly (RVE_para,RVEModel ,RVEPart):
    #Initial parameter
    RVEAssembly =RVEModel.rootAssembly
    #Create a independent mesh Instance
    RVEAssembly.Instance(name='', part=, dependent=)
    # Creat nine reference points, and set a set
    RVEAssembly.ReferencePoint(point=)
    RVEAssembly.Set(referencePoints=, name=)
    #Define two types of the mesh
    elemType1 = mesh.ElemType(elemCode=,elemLibrary=)
    elemType2 = mesh.ElemType(elemCode=,elemLibrary=)
    #Assign two types of the mesh to the fiber and matrix respectively
    RVEAssembly.setElementType(regions=, elemTypes=)
    #creat mesh seed
    RVEAssembly.seedEdgeBySize(edges=, size=)
    #mesh the partInstances
    RVEAssembly.generateMesh(regions=)
    return RVEAssembly
```

## 5.5. The displayed field (Create step)

This function runs the ABAQUS to creat the analysis step of the RVE. In this study, the outdata fields are strain ($E$), displacement ($U$), and stress ($S$).

```
##Create RVE step
def CreateStep (RVE_para,RVEModel ):
    #Define a static step
    RVEModel.StaticStep(name='', previous='')
    #Define the outdata
    RVEModel.fieldOutputRequests[''].setValuesInStep(stepName='',variables=)
```

## 5.6. Imposition of periodic boundary conditions (Create MPCs)

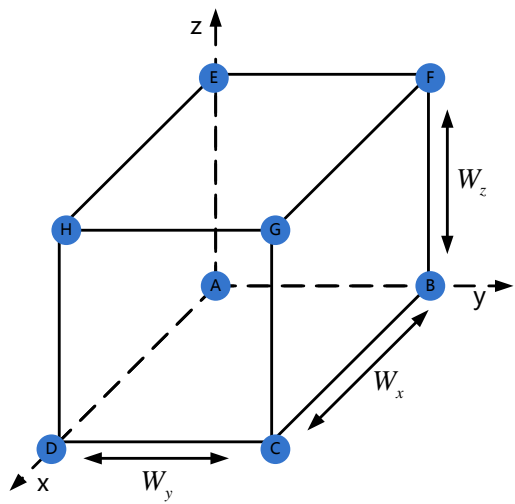

Fig.3 Prescribed PBC of 3-D case

There are three types of node sets: faces, edges and vertices. To avoid redundant constraints on edges and vertices, the nodes on edges and vertices are excluded from faces, and nodes on vertices are also excluded from edges when defining node sets in order to impose PBCs. A vertex node set contains only one node, while each node set for edges and faces could contain multiple nodes. The equations of all faces, edges, and vertices are as follows:

The faces:

$$\begin{cases} u_{HGCD} - u_{EFBA} = W_x \varepsilon_x^0 \\ v_{HGCD} - v_{EFBA} = 0 \\ w_{HGCD} - w_{EFBA} = 0 \end{cases} \begin{cases} u_{BCGF} - u_{ADHE} = 0 \\ v_{BCGF} - v_{ADHE} = W_y \varepsilon_y^0 \\ w_{BCGF} - w_{ADHE} = 0 \end{cases} \begin{cases} u_{EFGH} - u_{ABCD} = 0 \\ v_{EFGH} - v_{ABCD} = 0 \\ w_{EFGH} - w_{ABCD} = W_z \varepsilon_z^0 \end{cases} \quad (19)$$

The edges:

$$\begin{cases} u_{CG} - u_{AE} = W_x \varepsilon_x^0 + W_y \varepsilon_{xy}^0 \\ v_{CG} - v_{AE} = W_x \varepsilon_{yx}^0 + W_y \varepsilon_y^0 \\ w_{CG} - w_{AE} = W_x \varepsilon_{zx}^0 + W_y \varepsilon_{zy}^0 \end{cases} \begin{cases} u_{HG} - u_{AB} = W_x \varepsilon_x^0 + W_z \varepsilon_{xz}^0 \\ v_{HG} - v_{AB} = W_x \varepsilon_{yx}^0 + W_z \varepsilon_{yz}^0 \\ w_{HG} - w_{AB} = W_x \varepsilon_{zx}^0 + W_z \varepsilon_z^0 \end{cases} \begin{cases} u_{FG} - u_{AD} = W_y \varepsilon_{xy}^0 + W_z \gamma_{xz}^0 \\ v_{FG} - v_{AD} = W_y \varepsilon_y^0 + W_z \varepsilon_{yz}^0 \\ w_{FG} - w_{AD} = W_y \varepsilon_{zy}^0 + W_z \varepsilon_z^0 \end{cases} \quad (20)$$

$$\begin{cases} u_{HD} - u_{FB} = W_x \varepsilon_x^0 - W_y \varepsilon_{xy}^0 \\ v_{HD} - v_{FB} = W_x \varepsilon_{yx}^0 - W_y \varepsilon_y^0 \\ w_{HD} - w_{FB} = W_x \varepsilon_{zx}^0 - W_y \varepsilon_{zy}^0 \end{cases} \begin{cases} u_{DC} - u_{EF} = W_x \varepsilon_x^0 - W_z \varepsilon_{xz}^0 \\ v_{DC} - v_{EF} = W_x \varepsilon_{yx}^0 - W_z \varepsilon_{yz}^0 \\ w_{DC} - w_{EF} = W_x \varepsilon_{zx}^0 - W_z \varepsilon_z^0 \end{cases} \begin{cases} u_{BC} - u_{EH} = W_y \varepsilon_{xy}^0 - W_z \varepsilon_{xz}^0 \\ v_{BC} - v_{EH} = W_y \varepsilon_y^0 - W_z \varepsilon_{yz}^0 \\ w_{BC} - w_{EH} = W_y \varepsilon_{zy}^0 - W_z \varepsilon_z^0 \end{cases} \quad (21)$$

The vertices:

$$\begin{cases} u_G - u_A = W_x \varepsilon_x^0 + W_y \varepsilon_{xy}^0 + W_z \varepsilon_{xz}^0 \\ v_G - v_A = W_x \varepsilon_{yx}^0 + W_y \varepsilon_y^0 + W_z \varepsilon_{xz}^0 \\ w_G - w_A = W_x \varepsilon_{zx}^0 + W_y \varepsilon_{zy}^0 + W_z \varepsilon_z^0 \end{cases} \begin{cases} u_F - u_D = -W_x \varepsilon_x^0 + W_y \varepsilon_{xy}^0 + W_z \varepsilon_{xz}^0 \\ v_F - v_D = -W_x \varepsilon_{yx}^0 + W_y \varepsilon_y^0 + W_z \varepsilon_{xz}^0 \\ w_F - w_D = -W_x \varepsilon_{zx}^0 + W_y \varepsilon_{zy}^0 + W_z \varepsilon_z^0 \end{cases} \quad (22)$$

$$\begin{cases} u_H - u_B = W_x \varepsilon_x^0 - W_y \varepsilon_{xy}^0 + W_z \varepsilon_{xz}^0 \\ v_H - v_B = W_x \varepsilon_{yx}^0 - W_y \varepsilon_y^0 + W_z \varepsilon_{xz}^0 \\ w_H - w_B = W_x \varepsilon_{zx}^0 - W_y \varepsilon_{zy}^0 + W_z \varepsilon_z^0 \end{cases} \begin{cases} u_C - u_E = W_x \varepsilon_x^0 + W_y \varepsilon_{xy}^0 - W_z \varepsilon_{xz}^0 \\ v_C - v_E = W_x \varepsilon_{yx}^0 + W_y \varepsilon_y^0 - W_z \varepsilon_{xz}^0 \\ w_C - w_E = W_x \varepsilon_{zx}^0 + W_y \varepsilon_{zy}^0 - W_z \varepsilon_z^0 \end{cases} \quad (23)$$

In the terminology of ABAQUS, these equations have been introduced as 'EQUATION'. The following is the steps in applying PBC on a model [20]:

(1) Group the node sets on faces, edges, and vertices of the RVE. The *EQUATION keyword allows only those nodes on opposite faces which have matching coordinates. Hence, regular or identical mesh on each the opposing faces are required. In case of irregular mesh, the outermost layer of the RVE geometry need be remeshed such that this condition is satisfied.

(2) The prescribed loading (displacement) is applied on a reference point situated

outside the RVE domain.

```
##Create RVE boundary conditons
def CreateBoudary (RVE_para,RVEModel ,RVEPart,RVEAssembly):
   #Initial parameter
   #find nodes, define MPC
    RVEAssembly.Set(nodes=,name='')
    RVEAssembly.Set(nodes=,name='')
    RVEModel.Equation(name='',terms=)
```

## 5.7. Load case generation (Create load)

Making use of key DOFs, six macroscopic strains $(\varepsilon_x^0, \varepsilon_y^0, \varepsilon_z^0, \varepsilon_{xy}^0, \varepsilon_{xz}^0, \varepsilon_{yz}^0)$ can be prescribed in any way, individually or in any combination.

In the code, six macroscopic stresses are prescribed individually as six load cases within a single step of analysis, so that uniaxial strain states are obtained, as required in Eq. (23). The user has the option of generating an additional combined load case by specifying the macroscopic strain components involved. Thus, three stiffness tensors $(C_{12}, C_{13}, C_{23})$ are obtained by the combined load case.

```
##Create the load
def CreateLoad (RVE_para,RVEModel ,RVEAssembly):
    region1 = RVEAssembly.sets['']
    RVEModel.DisplacementBC(name='', createStepName='',region=,u1=)
```

## 5.8. Generating a analysis work (Create job)

This function runs the ABAQUS to creat the analytical task of the RVE. The display of the post-processing is set up, and the precision of the outdata is also set up in this function. To save computation time, the number of the CPU may be added to accelerate the computation.

```
##Create the job
def CreateJob (RVE_para):
    # Inital parameter
    RVEJob = mdb.Job(name=, model=)
    # Submint job
    RVEJob.submit()
    #Wait the job to complete
    RVEJob.waitForCompletion()
```

## 5.9. The computation of the stiffness tensor (Create post-processing)

This function outputs the analysis results of the RVE. With the macroscopic strains being expressed in terms of 'displacements' applied to RVEs, all effective properties of the RVEs are obtained in terms of the key DOFs. As the post-process of averaging stresses and strains are not usually available directly from commercial FE codes, the ABAQUS offers extremely useful means of simplifying the post-processing of FE analysis for RVEs.

```
def CreateResult (RVE_para,stif):
    # Inital parameter
    odb = openOdb(path=)
    history = odb.steps["].historyRegions["].historyOutputs["]
    subHistory = history.data
    # Averaged stress
    #Total energy
    RVEenergy = subHistory[len(subHistory)-1][1]
    RVEstress=2*RVEenergy/(RVEvolume*RVEstrain)
```

## 5.10. Main Program (Create the initial parameter)

The main program defines the global execution flow. The main program locates at the end of the source code since the Python interpreter needs the definition of all other functions before they are called in the main program [19].

In the current main program, this program is followed by the function body that can be partitioned into three blocks: input initialization, iteration loop, and compliance matrix.

(1) Input initialization block: the initial design parametric model is introduced in the following. Since the information in the design part, such as the strain information, is frequently used in the program. Thus, all parameters are predefined referring to the objects: the mesh seed, the length of the RVE, the fiber volume fraction, the type of the element, the material properties of the fiber and matrix, and the prescribed strain. A model database object is created referring to the input design model and assigned to a structure variable *RVE_para*.

(2) Iteration loop: for the stiffness tensor, there are nine unknown parameters in the matrix for the unidirectional composite. Hence, nine jobs are created to obtain the unknown stiffness tensor parameters as shown in Figure 4. Except for MPCs and Loads, other modules such as the geometry, material properties, and mesh are the same to be created in this study. The MPCs and Loads modules are created based on different load cases such as $(\varepsilon_x^0 = 1; \varepsilon_y^0, \varepsilon_z^0, \varepsilon_{xy}^0, \varepsilon_{xz}^0, \varepsilon_{yz}^0 = 0)$. The load case $(\varepsilon_x^0 = 1; \varepsilon_y^0, \varepsilon_z^0, \varepsilon_{xy}^0, \varepsilon_{xz}^0, \varepsilon_{yz}^0 = 0)$ is to compute the stiffness tensor $C_{11}$.

(3) Compliance matrix: the compliance tensor is obtained by the inverse of the stiffness tensor as shown in Eq. (24)

$$[S] = \begin{bmatrix} S_{11} & S_{12} & S_{13} & 0 & 0 & 0 \\ S_{12} & S_{22} & S_{23} & 0 & 0 & 0 \\ S_{13} & S_{23} & S_{33} & 0 & 0 & 0 \\ 0 & 0 & 0 & S_{44} & 0 & 0 \\ 0 & 0 & 0 & 0 & S_{55} & 0 \\ 0 & 0 & 0 & 0 & 0 & S_{66} \end{bmatrix} = \begin{bmatrix} C_{11} & C_{12} & C_{13} & 0 & 0 & 0 \\ C_{12} & C_{22} & C_{23} & 0 & 0 & 0 \\ C_{13} & C_{23} & C_{33} & 0 & 0 & 0 \\ 0 & 0 & 0 & C_{44} & 0 & 0 \\ 0 & 0 & 0 & 0 & C_{55} & 0 \\ 0 & 0 & 0 & 0 & 0 & C_{66} \end{bmatrix}^{-} \quad (24)$$

Thus, the effective properties can be obtained as follows from the analysis of the RVEs:

$$\begin{aligned} E_1 &= \frac{1}{S_{11}} & v_{12} &= -\frac{S_{12}}{S_{11}} & G_{12} &= \frac{1}{S_{44}} \\ E_2 &= \frac{1}{S_{22}} & v_{13} &= -\frac{S_{13}}{S_{11}} & G_{13} &= \frac{1}{S_{55}} \\ E_3 &= \frac{1}{S_{33}} & v_{23} &= -\frac{S_{23}}{S_{22}} & G_{23} &= \frac{1}{S_{66}} \end{aligned} \quad (25)$$

Due to the unidirectional boron/aluminum RVE, predicted elastic properties are $E_1 = E_2, G_{13} = G_{23}$ and $v_{13} = v_{23}$. Moreover, it is noted that nine indendent material constants must be determined for the RVE. Therefore, in this study, in total there are nine ABAQUS analytical tasks to obtain the nine indendent material constants. Three Poisson'ratios $v_{12}, v_{13},$ and $v_{23}$ are computed after the computation of $E_1, E_2,$ and $E_3$ as shown in Figure 4. To improve the computation time of obtaining each of the above properties, there is a parallel method to simultaneously calculate the multiple material constants. On the other hand, calculating them individually from the RVE analysis and finding out the relationships among material constants can serve as a valuable check on the RVE. In particular, the correct application of all the boundary conditions for the RVE is important and a more sophisticated study.

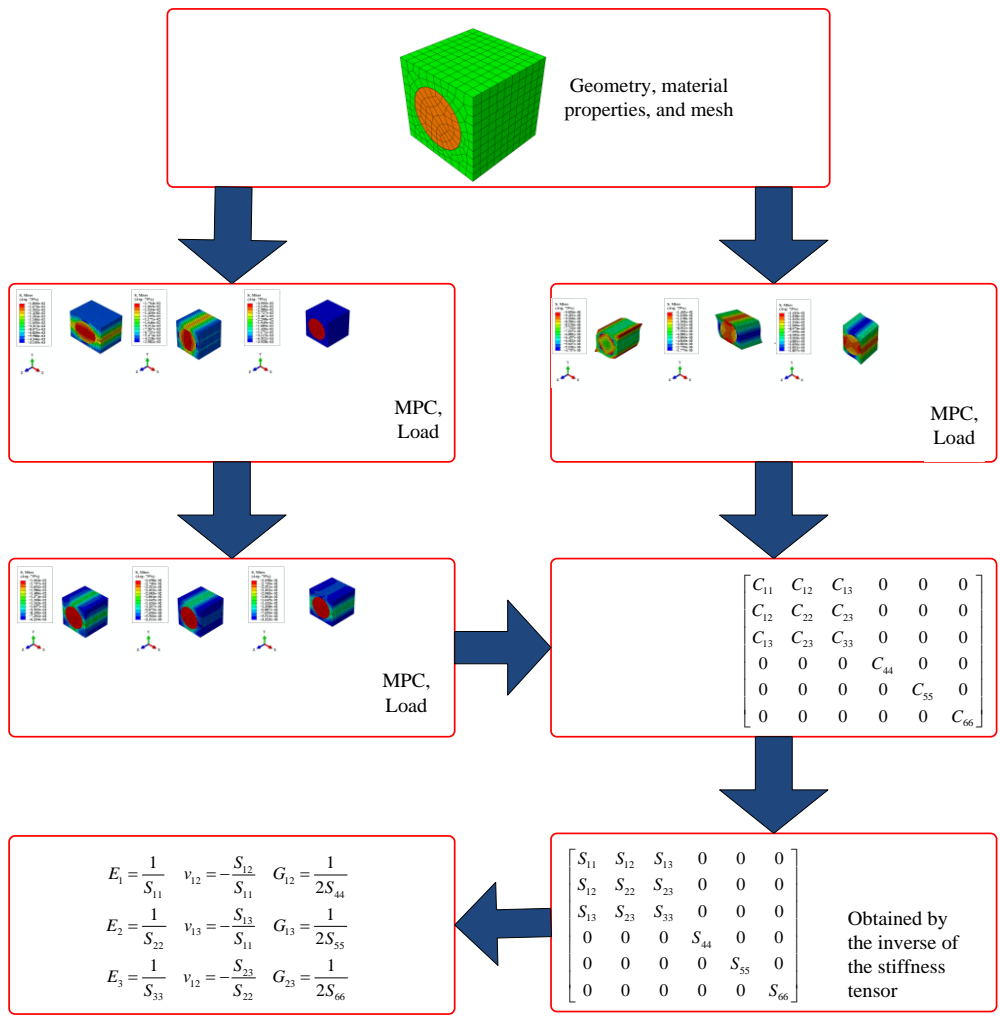

Fig. 4 The computation flow of the code

```
##========Main Program=======
##Set parameters and input
RVE_para={}
stif=zeros([6,6])
RVE_para["mesh"]=0.05
RVE_para["length"]=1.0000
RVE_para["volumeFraction"]=0.47
RVE_para["elementCode"]=C3D20
RVE_para["matrixProperties"]=(68.3,0.3)
RVE_para["fiberProperties"]=(379.3,0.1)
RVE_para["strain"]=0.0001
RVE_para["prestrain"]=[0.0000,0.0000,0.0000,0.0000,0.0000,0.0000]
for n in range(len(RVE_para["prestrain"])):
    RVE_para["prestrain"]=[0.0000,0.0000,0.0000,0.0000,0.0000,0.0000]
    RVE_para["number"]=n
    RVE_para["name"]='RVE'+str(n+1)
    RVEModel = mdb.Model(RVE_para["name"])
    if (n<6):
        RVE_para["prestrain"][n]=RVE_para["strain"]
    else:
        RVE_para["prestrain"]=[RVE_para["strain"],RVE_para["strain"],RVE_para["strain"],
                0.0000,0.0000,0.0000]
        k=8-n
        RVE_para["prestrain"][k]=0.0000
    RVEPart=CreatePart(RVE_para,RVEModel)
    CreateMaterial(RVE_para,RVEModel)
    CreateSection(RVE_para,RVEModel,RVEPart)
    RVEAssembly=CreateAssembly (RVE_para,RVEModel,RVEPart)
    CreateStep (RVE_para,RVEModel )
    CreateBoudary (RVE_para,RVEModel,RVEPart,RVEAssembly)
    CreateLoad (RVE_para,RVEModel,RVEAssembly)
    CreateJob (RVE_para)
    if (n<6):
        stif[n][n]=CreateResult(RVE_para,stif)
    if (5<n<8):
        stif[0][n-5]=CreateResult(RVE_para,stif)
        stif[n-5][0]=CreateResult(RVE_para,stif)
    if (n==8):
        stif[n-7][n-6]=CreateResult(RVE_para,stif)
        stif[n-6][n-7]=CreateResult(RVE_para,stif)
print stif, print "change"
stif1=mat(stif)
print stif1
compliance=stif1**(-1)
print "compliance matrix",print compliance
```

## 6. Results

The visualization module of ABAQUS can be used to output von Mises' stress as shown in Figure 5. It is observed that the deformations are generally similar to the results in the literature [18].

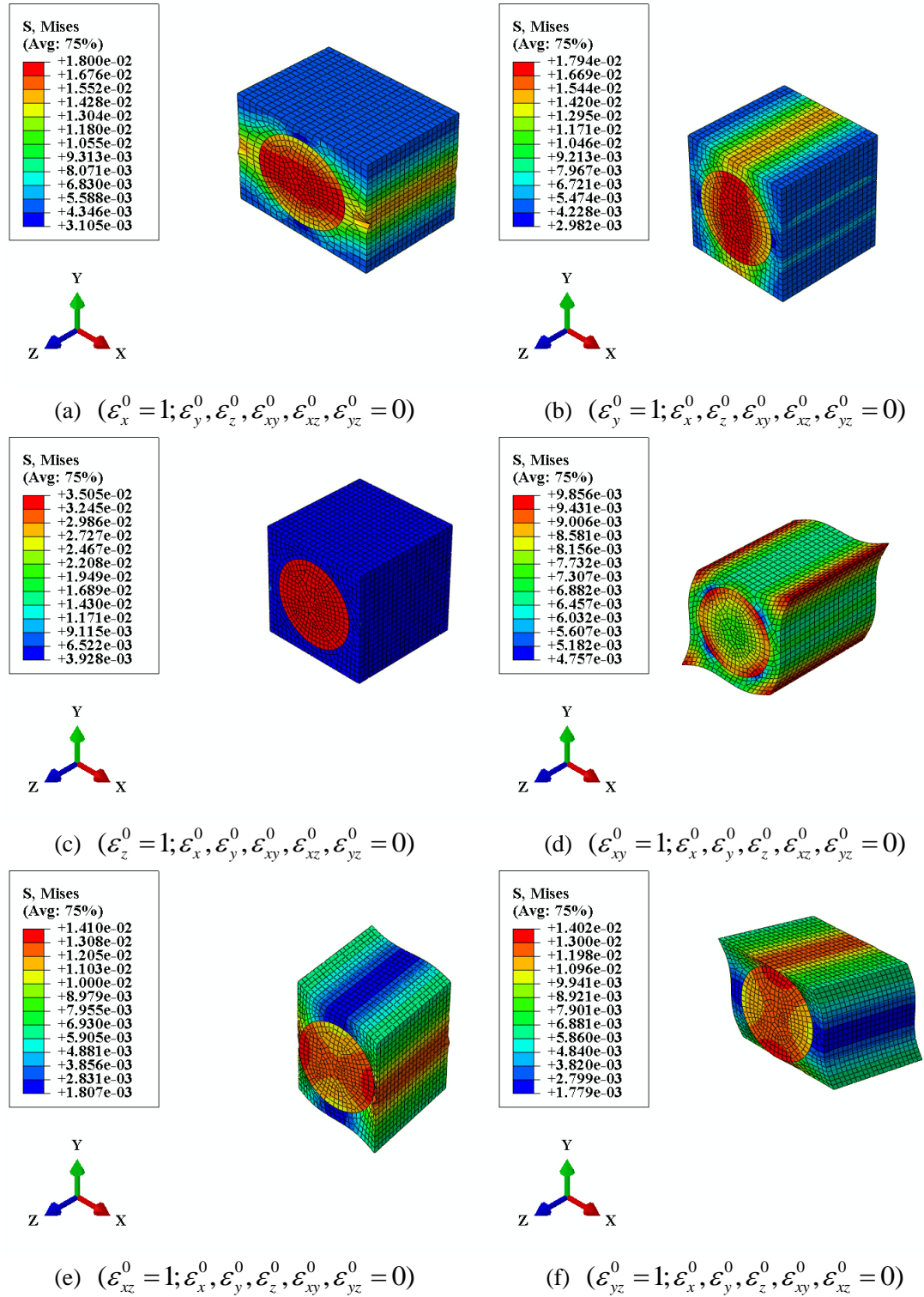

(a) $(\varepsilon_x^0 = 1; \varepsilon_y^0, \varepsilon_z^0, \varepsilon_{xy}^0, \varepsilon_{xz}^0, \varepsilon_{yz}^0 = 0)$

(b) $(\varepsilon_y^0 = 1; \varepsilon_x^0, \varepsilon_z^0, \varepsilon_{xy}^0, \varepsilon_{xz}^0, \varepsilon_{yz}^0 = 0)$

(c) $(\varepsilon_z^0 = 1; \varepsilon_x^0, \varepsilon_y^0, \varepsilon_{xy}^0, \varepsilon_{xz}^0, \varepsilon_{yz}^0 = 0)$

(d) $(\varepsilon_{xy}^0 = 1; \varepsilon_x^0, \varepsilon_y^0, \varepsilon_z^0, \varepsilon_{xz}^0, \varepsilon_{yz}^0 = 0)$

(e) $(\varepsilon_{xz}^0 = 1; \varepsilon_x^0, \varepsilon_y^0, \varepsilon_z^0, \varepsilon_{xy}^0, \varepsilon_{yz}^0 = 0)$

(f) $(\varepsilon_{yz}^0 = 1; \varepsilon_x^0, \varepsilon_y^0, \varepsilon_z^0, \varepsilon_{xy}^0, \varepsilon_{xz}^0 = 0)$

Figure 5 The von Mises' Stress field corresponding to different loads.

The orthotropic elastic matrix is obtained by applying PBC to the RVE. The results of the stiffness tensor are given as follows by using ABAQUS.

$$\begin{bmatrix} \varepsilon_{11} \\ \varepsilon_{22} \\ \varepsilon_{33} \\ \varepsilon_{12} \\ \varepsilon_{13} \\ \varepsilon_{23} \end{bmatrix} = \begin{bmatrix} 0.006962 & -0.001779 & -0.000906 & 0 & 0 & 0 \\ -0.001779 & 0.006962 & -0.000906 & 0 & 0 & 0 \\ -0.000906 & -0.000906 & 0.004653 & 0 & 0 & 0 \\ 0 & 0 & 0 & 0.021864 & 0 & 0 \\ 0 & 0 & 0 & 0 & 0.018437 & 0 \\ 0 & 0 & 0 & 0 & 0 & 0.018434 \end{bmatrix} \begin{bmatrix} \sigma_{11} \\ \sigma_{22} \\ \sigma_{33} \\ \sigma_{12} \\ \sigma_{13} \\ \sigma_{23} \end{bmatrix}$$

It is seen from examining Table 3 that the predicted properties are generally in good agreement with the results in the literature, and the experimental values.

Table 3 Results and comparison for unidirectional boron/aluminum RVE ($V_f$=0.47)[*]

| material constants | $E_3(Gpa)$ | $E_2(Gpa)$ | $G_{23}(Gpa)$ | $G_{12}(Gpa)$ | $V_{23}$ | $V_{12}$ |
|---|---|---|---|---|---|---|
| Present | 213 | 143 | 53.8 | 45.4 | 0.194 | 0.256 |
| Ref.13 | 214 | 143 | 54.2 | 45.7 | 0.195 | 0.253 |
| Ref.14 | 215 | 144 | 57.2 | 45.9 | 0.19 | 0.29 |
| Ref.21 | 214 | 135 | 51.1 | - | 0.19 | - |
| Ref.22 | 214 | 156 | 62.6 | 43.6 | 0.20 | 0.31 |
| Ref.23 | 215 | 123 | 53.9 | - | 0.19 | - |
| Ref.24 | 215 | 135.2 | 53.9 | 52.3 | 0.195 | 0.295 |
| Ref.25 | 216 | 140 | 52 | - | 0.29 | - |

## 7. Conclusions

This study presents a Python code for computing effective properties of the RVE under PBCs. With simple modifications, the basic Python code may be extended to the computation of the effective properties of the more complex microstructure such as several inclusions with the different shapes, different orientations and different aspect ratios, and even the random distribution of inclusions. More importantly, the code provides a convenient platform upon which further extensions such as the closed loop optimization for the material and geometric parameter design of the composite microstructure, and different optimizers could be also easily built. In doing so, the user may experiment with various algorithms and tackle a wide range of problems. The Python code is also presented for educational and engineering practice purposes. Moreover, with extensions such as those presented in this study, and with the ABAQUS FEA/modelling power, the Python code is capable of solving further engineering design problems by using efficient optimization methods.

---

[*] The principal direction of the RVE is $z$-direction in this study, and the principal direction of the RVE is $x$-direction in the references.

## Acknowledgements

This study was funded by Project of the Program of National Natural Science Foundation of China (NSFC) grant number 11572120 and 61232014; Program for New Century Excellent Talents in University under the grant number NCET-11-0131.

## Appendix

### 1. A two-dimensional case

A 2-D RVE model is created. The model consists of a fiber reinforcement and matrix, with a volume fraction of 50 %. The properties of the materials are given in Table 4.

Table 4 Material properties of fiber and matrix

| Material | $E$ (Gpa) | $v$ |
|---|---|---|
| Fiber | 72.5 | 0.22 |
| Matrix | 2.6 | 0.4 |

The unidirectional RVE is assumed to be orthotropic and linearly elastic. For 2D RVE, from Eq. (13), the constitutive relation of this effective material can be written as

$$[\bar{\sigma}] = [C][\bar{\varepsilon}] \quad (26)$$

where $[C]$ is the stiffness matrix.

$$[C] = \begin{bmatrix} C_{11} & C_{12} & 0 \\ C_{12} & C_{22} & 0 \\ 0 & 0 & C_{66} \end{bmatrix} \quad (27)$$

After obtaining $\bar{\sigma}_{ij}$ for given $\bar{\varepsilon}_{ij}$ from the computation of a RVE, $C_{ij}$ can be obtained from Eq. (26). The relation between the homogeneous strain energy and the average strain is

$$\frac{U}{V_{\text{RVE}}} = \frac{1}{2} C_{11} \bar{\varepsilon}_1^2 + C_{12} \bar{\varepsilon}_1 \bar{\varepsilon}_2 + \frac{1}{2} C_{22} \bar{\varepsilon}_2^2 + \frac{1}{2} C_{66} \bar{\varepsilon}_6^2 \quad (28)$$

## 2. Imposition of periodic boundary conditions

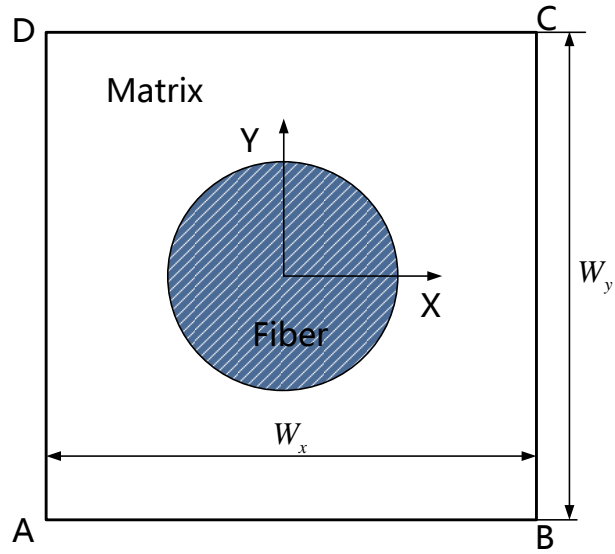

Fig. 6 Prescribed PBC of 2-D case

Periodic boundary conditions:

The edges:

$$\begin{cases} u_{BC} - u_{AD} = W_x \varepsilon_x^0 \\ v_{BC} - v_{AD} = W_x \varepsilon_{yx}^0 \end{cases} \begin{cases} u_{CD} - u_{AB} = W_y \varepsilon_{xy}^0 \\ v_{CD} - v_{AB} = W_y \varepsilon_y^0 \end{cases} \quad (29)$$

The vertices:

$$\begin{cases} u_C - u_A = W_x \varepsilon_x^0 + W_y \varepsilon_{xy}^0 \\ v_C - v_A = W_y \varepsilon_y^0 + W_x \varepsilon_{yx}^0 \end{cases} \begin{cases} u_B - u_D = W_x \varepsilon_x^0 - W_y \varepsilon_{xy}^0 \\ v_B - v_D = -W_y \varepsilon_y^0 + W_x \varepsilon_{yx}^0 \end{cases} \quad (30)$$

# 3. Python implement of the two-dimensional case

```
##========Main Program=======
##Set parameters and input
RVE_para={}
stif=zeros([3,3])
RVE_para["mesh"]=0.05
RVE_para["length"]=1.0000
RVE_para["volumeFraction"]=0.5
RVE_para["elementCode"]='CPS3'
RVE_para["matrixProperties"]=(2.6,0.4)
RVE_para["fiberProperties"]=(72.5,0.22)
RVE_para["strain"]=0.0001
RVE_para["prestrain"]=[0.0000,0.0000,0.0000]
for n in range(len(RVE_para["prestrain"])+1):
    RVE_para["prestrain"]=[0.0000,0.0000,0.0000]
    RVE_para["number"]=n
    RVE_para["name"]='RVE'+str(n+1)
    RVEModel = mdb.Model(RVE_para["name"])
    if (n<3):
        RVE_para["prestrain"][n]=RVE_para["strain"]
    else:
        RVE_para["prestrain"]=[RVE_para["strain"],RVE_para["strain"],0.0000]
    RVEPart=CreatePart(RVE_para,RVEModel)
    CreateMaterial(RVE_para,RVEModel)
    CreateSection(RVE_para,RVEModel,RVEPart)
    RVEAssembly=CreateAssembly (RVE_para,RVEModel,RVEPart)
    CreateStep (RVE_para,RVEModel )
    CreateBoudary (RVE_para,RVEModel,RVEPart,RVEAssembly)
    CreateLoad (RVE_para,RVEModel,RVEAssembly)
    CreateJob (RVE_para)
    if (n<3):
        stif[n][n]=CreateResult(RVE_para,stif)
    if (n==3):
        stif[n-3][n-2]=CreateResult(RVE_para,stif)
        stif[n-2][n-3]=CreateResult(RVE_para,stif)
print stif
print "change"
stif1=mat(stif)
print stif1
compliance=stif1.I
print "compliance matrix"
print compliance
```